\documentclass[conference]{IEEEtran}
\IEEEoverridecommandlockouts
\usepackage{cite}
\usepackage{amsmath,amssymb,amsfonts}
\usepackage{algorithmic}
\usepackage{graphicx}
\usepackage{epstopdf}
\usepackage{textcomp}
\usepackage{xcolor}
\usepackage{subfigure}
\usepackage{enumerate}
\usepackage{booktabs}
\usepackage{multirow}

\usepackage{float}

\def\BibTeX{{\rm B\kern-.05em{\sc i\kern-.025em b}\kern-.08em
    T\kern-.1667em\lower.7ex\hbox{E}\kern-.125emX}}

\makeatletter
\newcommand{\linebreakand}{%
  \end{@IEEEauthorhalign}
  \hfill\mbox{}\par
  \mbox{}\hfill\begin{@IEEEauthorhalign}
}
\makeatother

\begin{document}
\title{Trajectory Generation and Tracking based on Energy Minimization for a Four-Link Brachiation Robot}
\author
{\IEEEauthorblockN{Zishang Ji\textsuperscript{$\dag$}}
\IEEEauthorblockA{\textit{School of Mechatronical Engineering}\\
\textit{Beijing Institute of Technology}\\
Beijing, China \\
zishangji@bit.edu.cn}
\and
\IEEEauthorblockN{Xuanyu Zhang\textsuperscript{$\dag$}}
\IEEEauthorblockA{\textit{School of Mechatronical Engineering} \\
\textit{Beijing Institute of Technology}\\
Beijing, China \\
xuanyuzhang@bit.edu.cn}
\and
\IEEEauthorblockN{Xuanzhe Wang}
\IEEEauthorblockA{\textit{School of Mechatronical Engineering} \\
\textit{Beijing Institute of Technology}\\
Beijing, China \\
xuanzhewang@bit.edu.cn}
% \linebreakand
\and
\IEEEauthorblockN{Yan Huang\textsuperscript{*}}
\centerline~\IEEEauthorblockA{\textit{1. School of Mechantronical Engineering, Beijing Institute of Technology} \\
\IEEEauthorblockA{\textit{2. Beijing Advanced Innovation Center for Intelligent Robots and Systems, Beijing Institute of Technology}}
\IEEEauthorblockA{\textit{3. Key Laboratory of Biomimetic Robots and Systems, Ministry of Education}}
Beijing, China \\
yanhuang@bit.edu.cn}
\thanks{$\dag$ Equal contribution}
\thanks{* Corresponding author}
\thanks{This work was supported by the National Natural Science Foundation of China (No. 62073038) and Beijing Institute of Technology Research Funds for High-Level Talents.}
}
% ------------------------------------------
% \author{
% 	\IEEEauthorblockN{
% 		Zishang Ji\IEEEauthorrefmark{1}\IEEEauthorrefmark{2}, 
% 		Xuanyu Zhang\IEEEauthorrefmark{1}\IEEEauthorrefmark{2}, 
%         % \thanks{$\dag$ * equal contribution}
%         % \thanks{$\dag$ These authors contributed equally to this work and should be considered co-first authors.}
% 		Xuanzhe Wang\IEEEauthorrefmark{1}, 
% 		Yan Huang\IEEEauthorrefmark{1}\IEEEauthorrefmark{3}\IEEEauthorrefmark{4}
% 	\IEEEauthorblockA{\IEEEauthorrefmark{1}School of Mechatronical Engineering, Beijing Institute of Technology, Beijing, 100081, China}
%     % \IEEEauthorblockA{\IEEEauthorrefmark{3}Corresponding author:}
%     \IEEEauthorblockA{\IEEEauthorrefmark{4}Corresponding author: yanhuang@bit.edu.cn}
%     \IEEEauthorblockA{\IEEEauthorrefmark{2}These authors contributed equally to this work and should be considered co-first authors}
%     }
% } 
\maketitle 
\begin{abstract}
Aiming to mimic the brachiation locomotion of primates, we establish a brachiation robot model capable of swinging between different bars. The robot's design is based on a four-link underactuated structure. We propose an offline trajectory generator with optimization for minimizing energy consumption, which is implemented by direct collocation method to generate joint-space trajectories. We also propose a linear Model Predictive Control (MPC) algorithm as the feedback controller. The proposed MPC concurrently tracks both trajectories in joint space and Cartesian space. In simulation experiments, we analyzed the influence of lower-to-upper arm length ratio and swing time on the motion performance. The simulation results also demonstrate the robot has satisfied ability in trajectory tracking, obstacle avoidance and robustness.
\end{abstract}


\section{Introduction}
Brachiation is a type of locomotion utilized by primates to navigate through complex tree branches. It involves grasping and swinging on discontinuous and uneven surfaces, such as member bars. Therefore, brachiation is a challenging issue to investigate and replicate in robotics. The capability of brachiation robots to traverse intricate aerial environments endows them highly valuable for applications in hazardous and inaccessible areas where human presence could be risky. 

So far, existing research on brachiation robots can be categorized into two types based on whether the robot has a body structure. Robots with a body primarily store energy through body's swing-back before brachiation and utilize inertia to increase the distance of the brachiation \cite{b1,b3,b5}. However, incorporating a body adds complexity to the control system and significantly increases the robot's mass, imposing higher demands on the motors. Robots without a body have relatively simpler dynamics. Moreover, it is easier to study the motion strategies of arms when the robot does not have a body. Therefore, a large number of studies are focusing on limb-based brachiation robots without a body \cite{b6,b7,b8,b9,b10,b11,b12,b13,b14,b15}, and we also follow this way.

Fukuda et al. designed a two-link brachiation robot and employed a heuristic approach to generate motion trajectories as feedforward control \cite{b6,b7,b8}. They also incorporated proportional-derivative (PD) feedback control. Meghdari et al. discussed an optimal trajectory generation method using Pontryagin's minimum principle \cite{b9}, then designed PD and adaptive robust controllers to track the optimal trajectories. Yamakawa et al. focused on a 2-degree of freedom(DOF) robot with a hook-shaped grasper, generating simple rigid-body kinematic trajectories based on pendulum motion and tracking them using PD control \cite{b10}\cite{b11}.
Another robot, ``Tarzan'', capable of moving on a flexible cable, was presented in \cite{b12,b13,b14}. Researchers designed energy-optimal trajectories using the multiple-shooting method and employed the linear quadratic regulator (LQR) for trajectory tracking.
The simplest possible prototype of a brachiation robot named ``AcroMonk" \cite{b15} was designed. The robot's motion was achieved through various methods, such as model-based time-varying LQR, model-free PD control and reinforcement learning-based control policy.

However, the aforementioned research on brachiation robots are primarily based on a two-link structure, which means that the upper and lower arms are regarded as just one link. This structure lacks sufficient biomimicry and can not be used to analyze the dynamics and control of a robot with upper limb and lower limb. Moreover, the limited DOF makes it difficult for the robot to avoid obstacles that cannot be grasped during motion. Therefore, the present study proposes a four-link model of a brachiation robot. Thus we can study trajectory generation and tracking of a more biomimetic brachiation robot.

Most of the trajectory generation methods used in the aforementioned research face challenges in handling complex constraints or can only solve convex optimization problems. Moreover, the presented four-link model has more DOFs and higher system complexity, making simple model-free control or those only consider current state errors inadequate. In contrast, model predictive control (MPC) can take into account the future motion of the system to better cope with the complex underactuated systems [16]. Therefore, in this study, we use the direct collocation method to generate trajectory, then linearize the dynamics and kinematics to apply linear MPC as the trajectory tracking controller.

The main contributions of this study are as follows:
\begin{enumerate}[1.]
\item Designing a four-link brachiation robot model. The model can swing between discontinuous bars and avoid obstacles that cannot be grasped.
% \item Proposing an offline trajectory generation using the direct collocation method. It generate joint space trajectories by minimizing energy consumption, considering dynamics model, obstacle avoidance and other common constraints.
\item Proposing an offline trajectory generation approach using the direct collocation method. It can generate joint space trajectories with minimizing energy consumption.
\item Proposing a linear MPC-based trajectory tracking method. Using linearized dynamics and kinematics models, this method can track trajectories in both joint space and Cartesian space.
\item Studying the influence of lower-to-arm length ratio and swing time on energy consumption.
\end{enumerate}

The rest of the paper is organized as follows: Section II presents the modeling of the robot's dynamics. Section III introduces the trajectory generation method. The MPC-based trajectory tracking is designed in Section IV. Section V presents simulation results and parameter study. In Section VI, we conclude the paper and discuss future work.
\section{Robot Modeling}
\begin{figure}
\centering
\includegraphics[scale = 0.1]{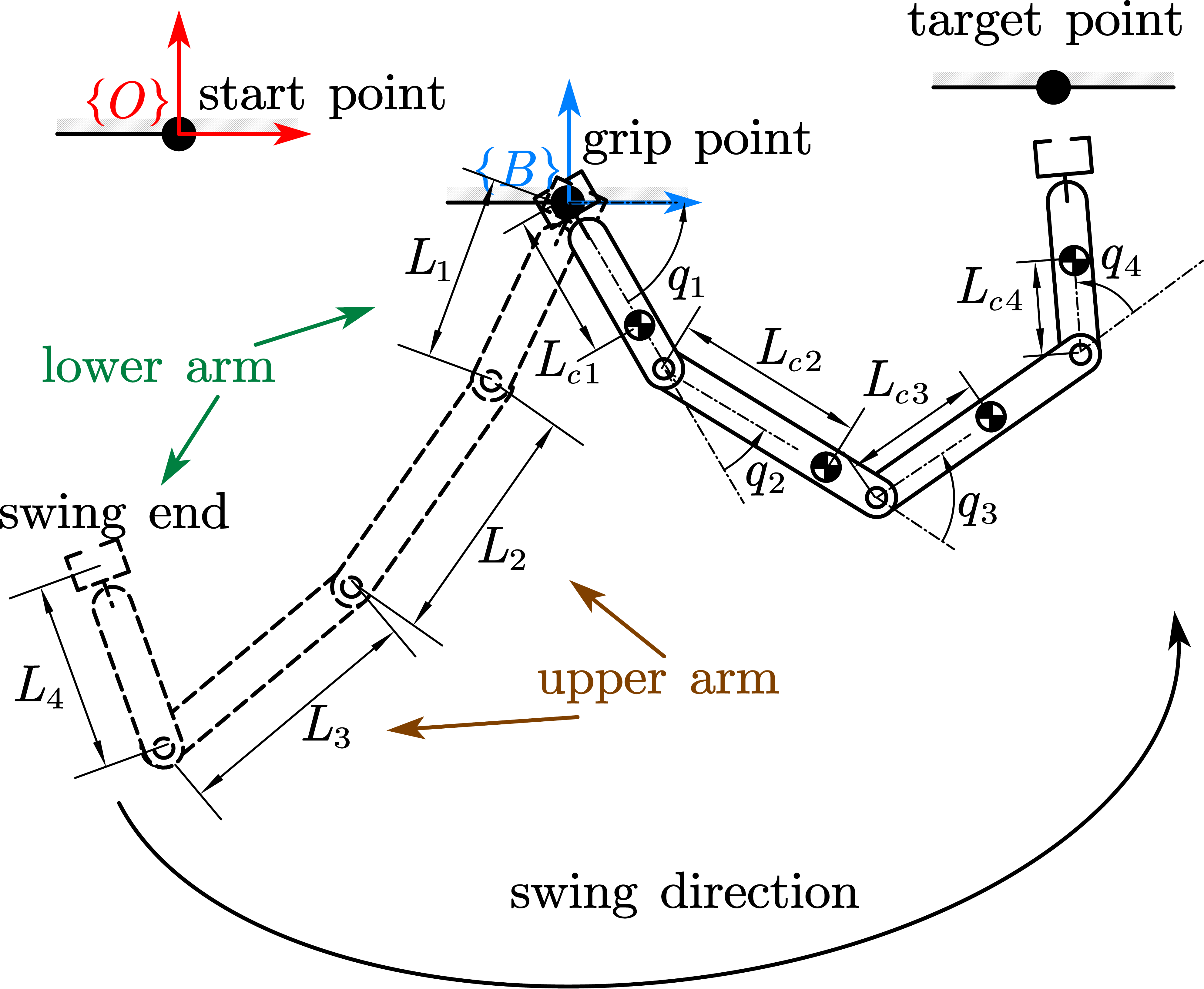}
\caption{Schematic diagram of the four-link brachiation robot's structure.}
\label{structure}
\end{figure}

Considering the real structure of the primates, we have developed a 4-link underactuated brachiation robot, depicted in Fig.~\ref{structure}. It comprises a pair of upper arms and lower arms, each terminating in a gripper. The underlying motion involves one arm grasping the bar at grip point while the other arms at swing end releases and swings towards the target point. The robot possesses four DOF, one of which is passive DOF at the gripper point ($q_1$). We fix the body frame $\left\{ B \right\} $ on the grip point, and the world frame $\left\{ O \right\} $ on the grip point of the first swing. The dynamics equation can be expressed as
\begin{align}
    M\left( \boldsymbol{q} \right) \ddot{\boldsymbol{q}}+C\left( \boldsymbol{q},\dot{\boldsymbol{q}} \right) \dot{\boldsymbol{q}}+G\left( \boldsymbol{q} \right) =B\boldsymbol{\tau },
\end{align}
where
\begin{align}
\nonumber
    \boldsymbol{q}=\left[ \begin{array}{c}
	q_1\\
	q_2\\
	q_3\\
	q_4\\
\end{array} \right] , \dot{\boldsymbol{q}}=\left[ \begin{array}{c}
	\dot{q}_1\\
	\dot{q}_2\\
	\dot{q}_3\\
	\dot{q}_4\\
\end{array} \right] , B= \begin{bmatrix}
	0&		0&		0\\
	1&		0&		0\\
	0&		1&		0\\
	0&		0&		1\\
\end{bmatrix} ,\boldsymbol{\tau }=\left[ \begin{array}{c}
	\tau _2\\
	\tau _3\\
	\tau _4\\
\end{array} \right]
\end{align}
in which $\boldsymbol{q}$ denotes the joint angles, $\boldsymbol{\tau}$ denotes the actuated joint torques. The dynamics equation is characterized by the mass matrix $M$, centrifugal and coriolis matrix $C$, gravity matrix $G$, and the selection matrix $B$.

For the sake of expediency in formulating the optimal control problem, the dynamic equations can be reformulated using a state-space representation:
\begin{align}
    \dot{\boldsymbol{x}}=f\left( \boldsymbol{x},\boldsymbol{u} \right)  = 
    \left[ \begin{array}{c}
	\dot{\boldsymbol{q}}\\
	M\left( \boldsymbol{q} \right) ^{-1}\left( B\boldsymbol{u}-G\left( \boldsymbol{q} \right) -C\left( \boldsymbol{q},\dot{\boldsymbol{q}} \right) \dot{\boldsymbol{q}} \right)\\
\end{array} \right], \label{state}
\end{align}
where $\boldsymbol{x}=\left[ \boldsymbol{q},\dot{\boldsymbol{q}} \right] ^T$ is the state vector, $\boldsymbol{u}=\boldsymbol{\tau }$ is the input vector.

\section{Trajectory Generation}
\begin{figure}
\centering
\includegraphics[scale = 0.24]{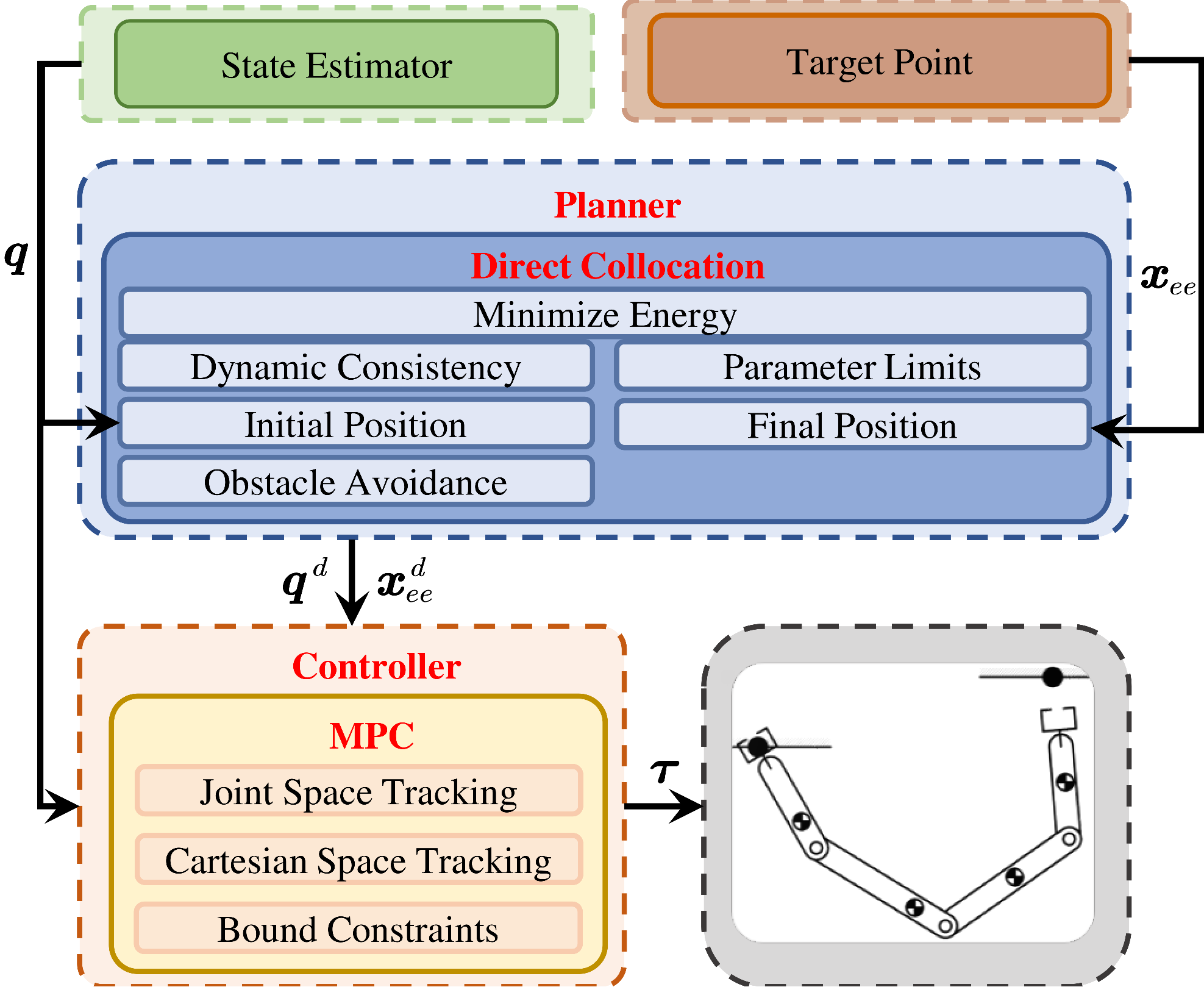}
\caption{Framework of planner and controller.}
\label{Control framework}
\end{figure}

Before each swing, our planner generates an offline trajectory that is tailored to the specific motion required. The control framework of the entire system is shown in Fig.~\ref{Control framework}. The nonlinear optimization framework we employ is structured as follows:
\begin{subequations}
\begin{align}
    \min_{\boldsymbol{q},\boldsymbol{\dot{q}},\boldsymbol{\tau}} \quad  &\text{Cost}(\boldsymbol{q},\boldsymbol{\dot{q}},\boldsymbol{\tau})  \label{opti:suba}\\ 
    \mathrm{s.t.} \quad    &\text{Dynamic Consistency}(\boldsymbol{q},\boldsymbol{\dot{q}},\boldsymbol{\tau})\label{opti:subb}\\ 
    &\text{Parameter Limits}(\boldsymbol{q},\boldsymbol{\dot{q}},\boldsymbol{\tau}) \label{opti:subc}\\
    &\text{Initial Position}(\boldsymbol{q},\boldsymbol{\dot{q}}) \label{opti:subd}\\ 
    &\text{Final Position}(\boldsymbol{q},\boldsymbol{\dot{q}}) \label{opti:sube}\\ 
    &\text{Obstacle Avoidance}(\boldsymbol{q}). \label{opti:subf} 
\end{align}
\end{subequations}
We solve this optimization problem using the \emph{direct collocation} method, which discretizing the original optimization framework over time, thereby transforming it into a large-scale numerical optimization problem. Given a known leaping time $T$, we discretize the trajectory into $N$ segments, resulting in $N+1$ mesh points. For brevity, we use the notation $\boldsymbol{x}_k\equiv \boldsymbol{x}\left( k \right) $ and $\boldsymbol{u}_k\equiv \boldsymbol{u}\left( k \right) $ to represent the state and control inputs at the k-th mesh point. Therefore, our decision variables consist of the state and control inputs at all mesh points:
\begin{equation}
\left\{ \boldsymbol{q}_1,\dot{\boldsymbol{q}}_1,\boldsymbol{\tau }_1,\cdots ,\boldsymbol{q}_{N+1},\dot{\boldsymbol{q}}_{N+1},\boldsymbol{\tau }_{N+1} \right\} .
\end{equation}
\subsection{Cost Function}\label{Cost Function}
Our goal is to make the most of gravity to minimize energy consumption and ensure that the trajectory is as smooth as possible, so the cost function\eqref{opti:suba} comprises two parts:
% \begin{equation}
%     \sum_{k=1}^{N+1}{\left( \dot{\boldsymbol{q}}_{k}^{T}Q_1\dot{\boldsymbol{q}}_k+\boldsymbol{\tau }_{k}^{T}R_1\boldsymbol{\tau }_k \right)}
% \end{equation}
\begin{equation}
    \sum_{k=1}^{N+1}{\left( \left\| \dot{\boldsymbol{q}}_k \right\| _{Q_1}^{2}+\left\| \boldsymbol{\tau }_k \right\| _{R_1}^{2} \right)},
\end{equation}
where $Q_1=Q_1^T\geqslant 0$ and $R_1=R_1^T\geqslant 0$ are weight matrices.

\subsection{Equality Constraints}\label{EqC}
The optimization framework includes three types of inequality constraints: \eqref{opti:subb} \eqref{opti:subd} and \eqref{opti:sube}. 
% Constraint \eqref{opti:subb} ensures that the entire trajectory satisfies the dynamic equations. Constraint \eqref{opti:subd} guarantees that the initial position is equal to the current measured position. Constraint \eqref{opti:sube} ensures that the robot can reach and grasp the target at the end.

\paragraph{Dynamics Consistency} We employ the dynamic equations to impose constraints between adjacent mesh points. These constraints are named as \emph{deft constraints} by Ferrolho\cite{b17}. It is more convenient to employ forward dynamic equations in underactuated system. At the k-th mesh point, the forward dynamic equations denoted as $f^{\mathrm{fd}}(\cdot)$ compute the acceleration $\boldsymbol{\ddot{q}_k}$ based on the current $\boldsymbol{q_k}$, $\boldsymbol{\dot{q}_k}$ and $\boldsymbol{\tau_k}$. By applying Euler's discretization method, we can obtain $\boldsymbol{q_{k+1}}$ and $\boldsymbol{\dot{q}_{k+1}}$, So equation \eqref{opti:subb} can be expressed as
\begin{align}
\nonumber \ddot{\boldsymbol{q}}_k=&f^{\mathrm{fd}}\left( \boldsymbol{q}_k,\dot{\boldsymbol{q}}_k,\boldsymbol{\tau }_k \right) 
\\
\dot{\boldsymbol{q}}_{k+1}=&\dot{\boldsymbol{q}}_k+\ddot{\boldsymbol{q}}_k\mathrm{dt}
\\
\nonumber \boldsymbol{q}_{k+1}=&\boldsymbol{q}_k+\frac{1}{2}\ddot{\boldsymbol{q}}_k\left( \mathrm{dt} \right) ^2,
\label{dyn_dis}
\end{align}
% where 应该是大写还是小写
where $\mathrm{dt}={{T}/{N}}$. It is important to note that with $N$ segments in the trajectory, the number of deft constraints is equal to $N$. Furthermore, we explicitly enforce $\boldsymbol{\tau_{N+1}}$ is equal to 0 in advance.

\paragraph{Initial Position} It is necessary to ensure alignment between the starting point of the robot's trajectory and the current state $q^{*}$ and $v^{*}$, so equation \eqref{opti:subd} can be formulated as
\begin{align}
\nonumber \boldsymbol{q}_1=&\boldsymbol{q}^*
\\
\dot{\boldsymbol{q}}_1=&\dot{\boldsymbol{q}}^*.
\end{align}
\paragraph{Final Position} To ensure that the robot end-effector(EE) can grasp the target at the end, we need to impose position constraints at the final time. As the given target is specified in Cartesian space $\boldsymbol{x}_{ee}$ and the 4-DOF robot has multiple solutions, the constraint \eqref{opti:sube} needs to be accomplished through the forward kinematic function $f^{\mathrm{fk}}(\cdot)$ instead of directly constraining the joint positions $\boldsymbol{q_{N+1}}$:
\begin{align}
    f^{\mathrm{fk}}\left( \boldsymbol{q_{N+1}} \right) =\boldsymbol{x}_{ee}
.\end{align}

Additionally, aiming to avoid any sudden collision, we impose a constraint that sets the EE's velocity to zero at the final time, so equation \eqref{opti:sube} also contains
\begin{align}
    J\left( \boldsymbol{q}_{N+1} \right) \cdot \dot{\boldsymbol{q}}_{N+1}=\mathbf{0}
.\end{align}
where $J$ represents the Jacobian matrix from the robot's body frame to the EE frame.
\subsection{Inequality Constraints}
The optimization framework includes two types of inequality constraints: \eqref{opti:subc} and \eqref{opti:subf}.
% Constraint \eqref{opti:subc} is used to restrict the parameters of the robot's joints, while constraint \eqref{opti:subf} ensures that the robot can avoid collisions with obstacles during its motion.

\paragraph{Parameter Limits} We implement simple \emph{boundary constraints} to restrict the parameters, so constraint \eqref{opti:subc} applies is
\begin{align}
    \nonumber \boldsymbol{q}_{\min}\leqslant \boldsymbol{q}_k\leqslant \boldsymbol{q}_{\max}
    \\
    \dot{\boldsymbol{q}}_{\min}\leqslant \dot{\boldsymbol{q}}_k\leqslant \dot{\boldsymbol{q}}_{\max}
    \\
    \nonumber \boldsymbol{\tau }_{\min}\leqslant \boldsymbol{\tau }_k\leqslant \boldsymbol{\tau }_{\max}.
\end{align}
\paragraph{Obstacle Avoidance} Assuming the presence of obstacles is known during the motion, we aim to plan a trajectory that avoids these obstacles. In trajectory optimization field, a conventional approach for obstacle avoidance is placing spherical collision primitives(CP) at critical parts of the robot and the obstacles, ensuring that the distances between these CPs exceed a specific threshold. However, for linked robots like ours, assigning multiple spherical CPs to 4 single links can lead to high computational costs. 
Zimmermann\cite{b18} introduced several common CPs in a unified form, among which the capsule CP is particularly suitable for linked robots. For each link $i$, a capsule CP can be placed. In our robot, the formulation of CP $i$ is given by
\begin{align}
    \mathbf{P}_i\left( q \right) =\mathbf{p}_i\left( q \right) +t\mathbf{v}_i\left( q \right),
\end{align}
where $\mathbf{P}_i(q)$ describes the coordinates of all points on CP $i$, $\mathbf{p}_i(q)$ and $\mathbf{v}_i(q)$ represent the start point and direction vector of the CP for link $i$, respectively. $t \in [0,1]$ is a scaling vector. We set the start point of capsule CP for link $i$ at joint $i$, the direction vector points from joint $i$ to joint $i + 1$. For the CP associated with obstacles, we use sphere CP, denoted as $\mathbf{P}_o = \mathbf{p}_o$ representing the cartesian coordinates of the obstacle.

The minimum distance between two CPs $A$ and $B$ can be calculated as
\begin{align}
    \mathcal{D} ^{AB}=\min_{0\leqslant t\leqslant 1} \left\| \mathbf{P}_A-\mathbf{P}_B \right\| ^2.
\end{align}

This raises a problem of finding the extremum of $ \mathcal{D} ^{AB}$. We can obtain the analytical solution or use optimization methods to find it. Finally, the constraint \eqref{opti:subf} is 
\begin{align}
    \mathcal{D}^{ij}\geqslant d_{\min}, \forall i,j\in \left\{ \mathrm{Link}_{1,2,3,4},\mathrm{Obstacle} \right\}.\label{D^{ij}}
\end{align}
where $d_{\min}$ is thresholds set in advance, which is usually the outer circle diameter of the obstacle. Equation \eqref{D^{ij}} serves to not only avoid collisions between the robot and obstacles in the environment, but also self-collisions.

\section{Trajectory Tracking}
In practical control, directly using the optimal control input $\boldsymbol{\tau}$ from offline trajectory generation can lead to error accumulation. Therefore, it is necessary to design a feedback controller to track the trajectory. In underactuated systems, the motion of the actuated joints significantly affects the states of the underactuated joints. Thus, in the current control cycle, when calculating the control input based on feedback and reference values, it is necessary to determine multiple control inputs for future time intervals to enable the robot to track the trajectory over a period of time in the future, especially for underactuated joints. To achieve this, MPC is employed. Considering the time sensitivity of underactuated systems, linear MPC is used to improve computational efficiency,.

We can start by performing a Taylor's formula of the right-hand side of equation \eqref{state} to obtain a linearized state equation around the point $\left( \boldsymbol{x}^*,\boldsymbol{u}^* \right)$, where $*$ denotes the measured states from state estimator:
\begin{align}
\dot{\boldsymbol{x}}=A\boldsymbol{x}+B\boldsymbol{u}+F,
\end{align}
where
\begin{align*}
    \nonumber A&=\left. \frac{\partial f\left( \boldsymbol{x},\boldsymbol{u} \right)}{\partial \boldsymbol{x}} \right|_{_{\boldsymbol{u}=\boldsymbol{u}^*}^{\boldsymbol{x}=\boldsymbol{x}^*}},B=\left. \frac{\partial f\left( \boldsymbol{x},\boldsymbol{u} \right)}{\partial \boldsymbol{u}} \right|_{_{\boldsymbol{u}=\boldsymbol{u}^*}^{\boldsymbol{x}=\boldsymbol{x}^*}},\\
	F&=f\left( \boldsymbol{x}^*,\boldsymbol{u}^* \right) -A\boldsymbol{x}^*-B\boldsymbol{u}^*.
\end{align*}

By using the forward Euler method to discretize the differential equation, we can obtain
\begin{align}
    \boldsymbol{x}\left( k+1 \right) =\bar{A}\boldsymbol{x}\left( k \right) +\bar{B}\boldsymbol{u}\left( k \right) +\bar{F}. \label{x(t)}
\end{align}

Our objective is to track the trajectory in both joint space and Cartesian space. Because the former will ensure the consistency of the motion pattern, and the latter will ensure the EE can accurately grasp the target. To track the trajectory in Cartesian space via linear MPC, we need to establish a linear output equation that relates the joint space to the Cartesian space. This can be achieved by using the forward kinematic equation $\boldsymbol{y}=f^{\mathrm{fk}}(\boldsymbol{q})$, where $\boldsymbol{y} = \begin{bmatrix} x_{ee} \ y_{ee} \end{bmatrix}^T$ denotes the Cartesian coordinates of the EE. By performing a Taylor's formula, similar to the state equation, we can get
\begin{align}
\boldsymbol{y}\left( k \right) =C\boldsymbol{x}\left( k \right) +f,  \label{y(t)}
\end{align}
where
{
\begin{align*}
C=\left[ \begin{matrix}
	\left. \frac{\partial x_{ee}}{\partial \boldsymbol{q}} \right|_{\boldsymbol{q}^*}&		0_{1\times 4}\\
	\left. \frac{\partial y_{ee}}{\partial \boldsymbol{q}} \right|_{\boldsymbol{q}^*}&		0_{1\times 4}\\
\end{matrix} \right] ,f=\left[ \begin{array}{c}
	x_{ee}\left( \boldsymbol{q}^* \right) -\left. \frac{\partial x_{ee}}{\partial \boldsymbol{q}} \right|_{\boldsymbol{q}^*}\boldsymbol{q}^*\\
	y_{ee}\left( \boldsymbol{q}^* \right) -\left. \frac{\partial y_{ee}}{\partial \boldsymbol{q}} \right|_{\boldsymbol{q}^*}\boldsymbol{q}^*\\
\end{array} \right] 
\end{align*}
}

For brevity, we use the notation $\boldsymbol{x}_k\equiv \boldsymbol{x}\left( k \right) $ and $\boldsymbol{u}_k\equiv \boldsymbol{u}\left( k \right) $. In the $k$-th control cycle, $\boldsymbol{x}_{k+1}$ can be found from $\boldsymbol{x}_{k}$ by equation \eqref{x(t)}. Subsequently, $\boldsymbol{y}_{k+1}$ can be found by equation \eqref{y(t)}. By analogy, both $\boldsymbol{x}$ and $\boldsymbol{y}$ can be found over the entire prediction horizon. Therefore, the objective function is
% \begin{align}
%     \nonumber
%     \underset{\boldsymbol{u}}{\min}\quad J\left( k \right) =\left( \boldsymbol{X}_k-{\boldsymbol{X}_k}^d \right) ^TQ\left( \boldsymbol{X}_k-{\boldsymbol{X}_k}^d \right) 
%     \\
%     +\left( \boldsymbol{Y}_k-{\boldsymbol{Y}_k}^d \right) ^TW\left( \boldsymbol{Y}_k-{\boldsymbol{Y}_k}^d \right) +{\boldsymbol{U}_k}^TR\boldsymbol{U}_k
% \end{align}
\begin{align}
    \nonumber
    \underset{\boldsymbol{U}_k}{\min}\,\,&J=\left\| \boldsymbol{X}_k-{\boldsymbol{X}_k}^d \right\| _{Q_2}^{2}+\left\| \boldsymbol{Y}_k-{\boldsymbol{Y}_k}^d \right\| _{W}^{2}+\left\| \boldsymbol{U}_k \right\| _{R_2}^{2} \\
    \mathrm{s}.\mathrm{t}.\,\, &\boldsymbol{U}_{\min}\leqslant \boldsymbol{U}_k\leqslant \boldsymbol{U}_{\max},
\end{align}
where
\begin{align*}
X_k=\left[ \begin{array}{c}
	x_{k+1}\\
	\vdots\\
	x_{k+N}\\
\end{array} \right] ,Y_k=\left[ \begin{array}{c}
	y_{k+1}\\
	\vdots\\
	y_{k+N}\\
\end{array} \right] ,U_k=\left[ \begin{array}{c}
	u_k\\
	\vdots\\
	u_{k+N-1}\\
\end{array} \right] 
\end{align*}
in which $N$ is the prediction horizon. The superscript $(\cdot)^d$ denotes the desired states obtained from the planner. ${X_k}^r$ and ${Y_k}^r$ has the same formula structure as $X_k$ and $Y_k$. $Q_2=Q_2^T\geqslant 0$,$W=W^T\geqslant 0$ and $R_2=R_2^T\geqslant 0$ are weight matrices.

To simplify the computation, the objective function can be rearranged into a quadratic planning(QP) form.
When applying control inputs, a receding-horizon approach is employed. 
% This approach involves solving the QP to obtain control inputs for the next $N$ time steps. However, it should be noted that only the control input for the first time step is applied to the joints.

\section{Experiments}

All experiments are conducted in simulation. The simulation platform is Webots. The direct collocation and QP in MPC is solved by open-source SNOPT and Quadprog++ library, respectively. To reduce the computation time of direct collocation, the time interval in equation \eqref{dyn_dis} is set to 10ms, then the trajectory points with a time interval of 1ms are generated by cubic spline interpolation and sent to the controller. The solution time for the direct collocation is within 200ms, while the solution time for MPC is within 1ms. The overall control cycle of the system is set to 1kHz.

\subsection{Parameter Study}\label{ps}
\begin{figure}[htbp]
\centering
\includegraphics[scale=0.12]{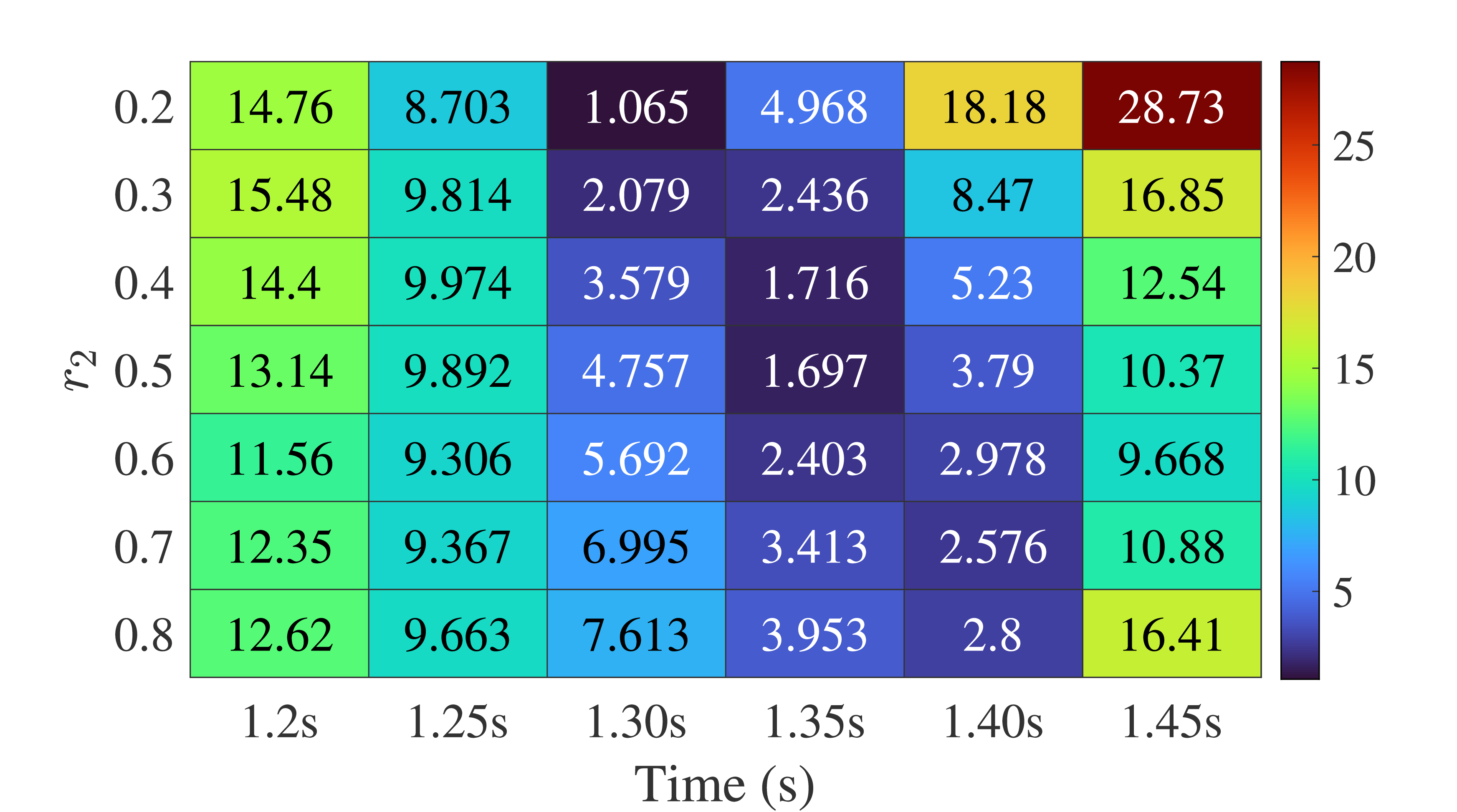}
\caption{Energy consumption (unit: $J$) across different $r_2$ and swing time.}
\label{Heatmap}
\end{figure}

\begin{figure}[htbp]
\centering
\subfigure[]
{
 	\begin{minipage}[b]{.9 \linewidth}
        \centering
        \includegraphics[scale=0.29]{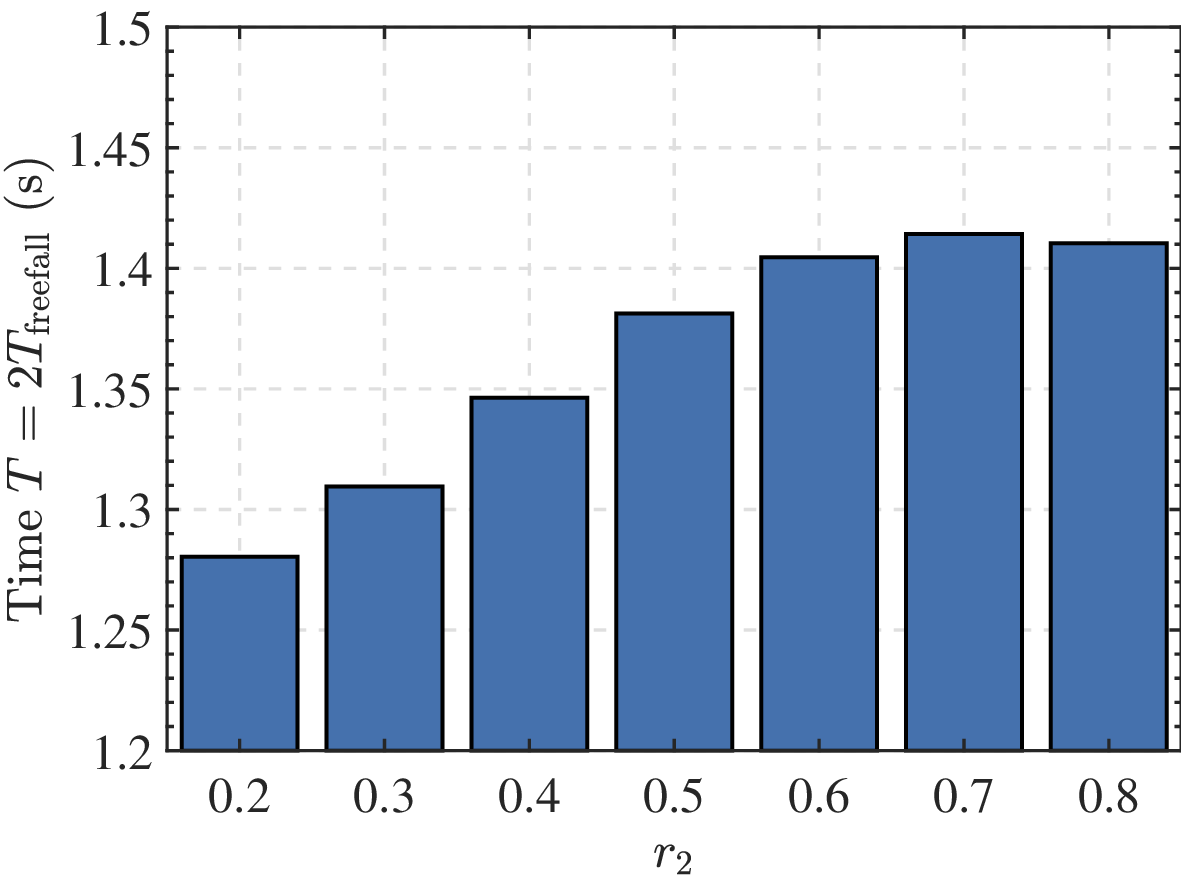}
    \end{minipage}
    \label{time_diff_R}
}
\subfigure[]
{
 	\begin{minipage}[b]{.9 \linewidth}
        \centering
        \includegraphics[scale=0.29]{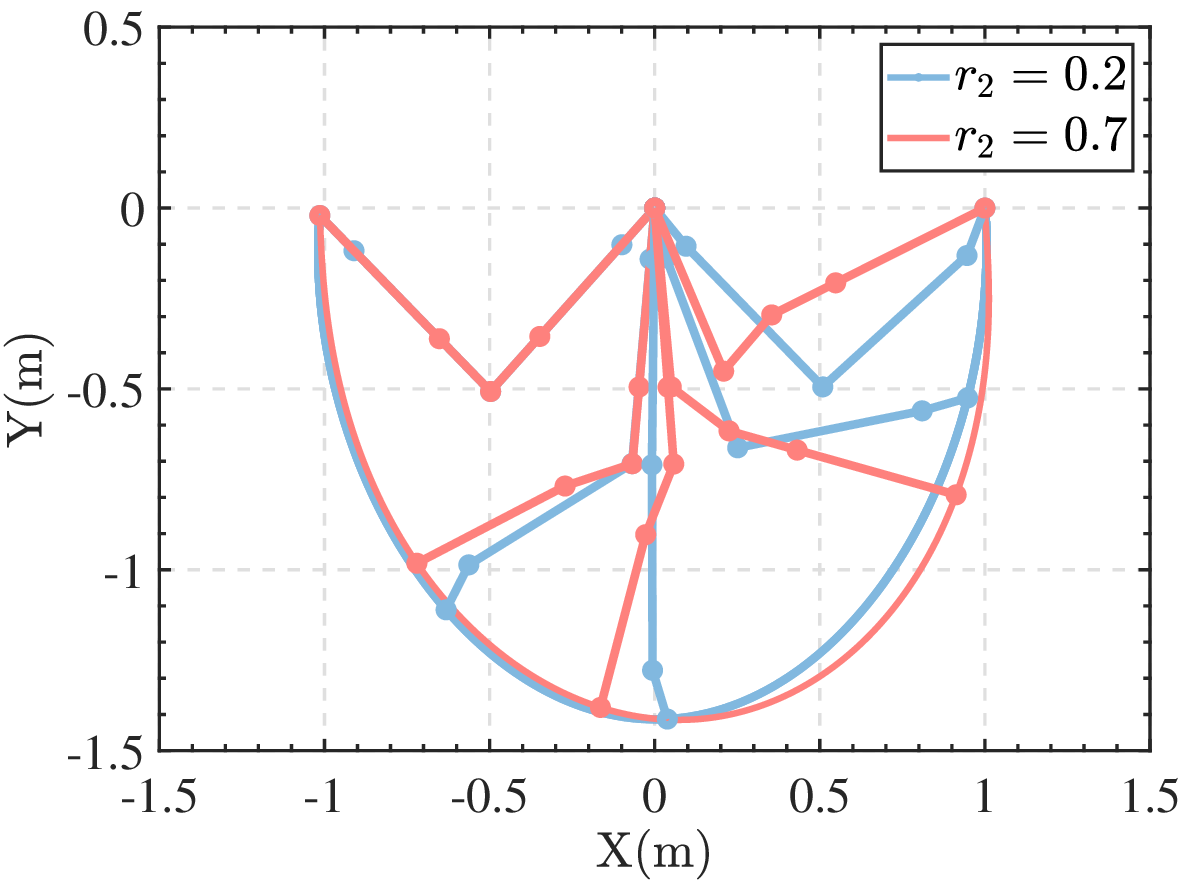}
    \end{minipage}
    \label{traj_diff_R}
}
\caption{\label{fig_diff_R} Swing characteristics at different $r_2$. (a) shows the time $T = 2T_{\text{freefall}}$ in different $r_2$; (b) shows two representative stick diagram of motion.}
\end{figure}

In the planner, we set the swing time $T = 2T_{\text{freefall}}$, where $T_{\text{freefall}}$ is the duration for the robot to undergo freefall motion from the start point to the horizontal midpoint between the start and target points. Drawing insights from bionics, we strive to incorporate physical parameter values of primates in real life which can be found in \cite{b19} into the selection of optimal parameters. While maintaining a constant total arm length 0.71m, we explore the influence of the lower-to-upper arm length ratio $R$ on energy consumption during swing, where $r_1 = L_1 / L_2 = L_4 / L_3$. For the convenience of expression below, we also define the ratio of the lower-to-total arm length ratio as $r_2 = L_1 / (L_1+L_2)$. Moreover, to verify the appropriateness of the swing time we set, we examine the different motion time as well. In simulation, we set the start point (-1m,0m) and target point ( 1m,0m). The positions of these two points are representative, because they make the swing distance not far or close. The calculation of energy consumption is 
\begin{align}
    E=\sum_{i=2}^4{\left( \int\limits_0^T{\dot{q}_i\cdot \tau _i\mathrm{dt}} \right)}.
\end{align}
Results are illustrated in Fig.~\ref{Heatmap} and Fig.~\ref{fig_diff_R}. We can draw the following three conclusions:
\begin{itemize}
    \item The minimum energy consumption is within the swing time range of $\left[ T-0.5s,T+0.5s \right]$, which provides validation for the rationality of our chosen swing time. Conversely, higher energy consumption during other time intervals can be attributed to the deviation of the robot's motion from the expected ``arc-like" swinging pattern. 
    \item As the length of the lower arm increases, the duration of motion required to achieve the minimum energy consumption also increases. This is because of the increase in $T_{\text{freefall}}$ shown in Fig.~\ref{time_diff_R}. Another reason is the change in the motion patterns. As depicted in Fig.~\ref{traj_diff_R}, when the lower arm is relatively short ($r_2 < 0.5 $), the robot just requires minimal effort to raise the lower arm at swing end during the latter half of the swing. In contrast, when the lower arm is longer ($r_2 \geqslant 0.5$), the robot necessitates more force to lift the lower arm. Remarkably, the robot primarily adjusts the joints near the grip point to ensure optimal utilization of inertia.
    \item The shorter the lower arm length, the smaller energy consumption during swing.
\end{itemize}

Considering the small difference in energy consumption between $r_2 = 0.5$ (closest to biological features) and $r_2 = 0.2 $ (with the lowest energy consumption), and the fact that robots with similar lengths between lower and upper arms exhibit better obstacle avoidance capabilities, we have chosen $r_2 = 0.5$ for our robot design. Another reason is that, from Fig.~\ref{Heatmap}, there is more tolerant of swing time errors when $r_2=0.5$. All parameters are summarized in Table~\ref{table}.
%---------------------------------------------------
% \begin{figure}[htbp]
% \raggedright
% \begin{minipage}[b]{.9\linewidth}
%         \includegraphics[scale=0.35]{EXP1_fig1_a.eps}
% \end{minipage}
% \vspace{-4mm}
% \caption{nameofFig. 3}
% \end{figure}
% \vspace{-.6cm}
% \begin{figure}[htbp]
% \centering
% \subfigure[]
% {
%  	\begin{minipage}[b]{.459 \linewidth}
%         \raggedleft
%         \includegraphics[scale=0.217]{EXP1_fig1_b_new.eps}
%     \end{minipage}
% }
% \subfigure[]
% {
%  	\begin{minipage}[b]{.459 \linewidth}
%         \raggedright
%         \includegraphics[scale=0.217]{EXP1_fig1_c.eps}
%     \end{minipage}
% }
% \caption{nameoffig4}
% \end{figure}

%---------------------------------------------------

%---------------------------------------------------
% \begin{figure}[htbp]
% \centering
% \subfigure[]
% {
%  	\begin{minipage}[b]{.459 \linewidth}
%         \raggedleft
%         \includegraphics[scale=0.215]{EXP1_fig1_b_new.eps}
%     \end{minipage}
%     \label{time_diff_R}
% }
% \subfigure[]
% {
%  	\begin{minipage}[b]{.459 \linewidth}
%         \raggedright
%         \includegraphics[scale=0.215]{EXP1_fig1_c.eps}
%     \end{minipage}
%     \label{traj_diff_R}
% }
% \caption{\label{fig_diff_R} Swing characteristics at different $r_2$. (a) shows the time that should have been set in the planner, (b) shows the representative trajectory.}
% \end{figure}
%---------------------------------------------------

\begin{table}[]
\caption{physical parameters of the simulated robot}
\centering
\begin{tabular}{ccc}
\hline
\textbf{Parameter}     & \textbf{Symbol}         & \textbf{Value}           \\ \hline
Link1 length       & $L_1$         & 0.355$m$     \\ 
Link2 length       & $L_2$         & 0.355$m$     \\ 
Link3 length       & $L_3$         & 0.355$m$     \\ 
Link4 length       & $L_4$         & 0.355$m$     \\ 
Link1 mass         & $m_1$         & 0.35$kg$     \\ 
Link2 mass         & $m_2$         & 0.56$kg$     \\ 
Link3 mass         & $m_3$         & 0.56$kg$     \\ 
Link4 mass         & $m_4$         & 0.35$kg$     \\ 
Link1 inertia   & $\left[I_{xx1},I_{yy1},I_{zz1}  \right] $ & {[}0.97 0.015  0.98{]}$kg\cdot m^2$   \\ 
Link2 inertia   & $\left[I_{xx2},I_{yy2},I_{zz2}  \right] $ & {[}0.10 0.024 0.10{]}$kg\cdot m^2$   \\ 
Link3 inertia   & $\left[I_{xx3},I_{yy3},I_{zz3}  \right] $ & {[}0.10 0.024 0.10{]}$kg\cdot m^2$   \\ 
Link4 inertia   & $\left[I_{xx4},I_{yy4},I_{zz4}  \right] $ & {[}0.97 0.015  0.98{]}$kg\cdot m^2$   \\ 
Link1 COM & $L_{com1}$     & 0.2059$m$            \\ 
Link2 COM & $L_{com2}$     & 0.1832$m$            \\ 
Link3 COM & $L_{com3}$     & 0.1718$m$            \\ 
Link4 COM & $L_{com4}$     & 0.1491$m$            \\
Max. torque & $\boldsymbol{\tau }_{\max}$     & {[}+5 +5  +5{]}$Nm$            \\
Min. torque & $\boldsymbol{\tau }_{\min}$     & {[}-5 -5  -5{]}$Nm$            \\
% \multirow{5}*{\shortstack{Diagonal in \\ weight matrix}}  & $Q_1$  &{[}1 1 1{]} \\
%                   & $R_1$  &{[}10 10 10{]} \\
%                   & $Q_2$  &{[}10 10 10 10 1 1 1 1{]} \\
%                   & $W$  &{[}$10^{-6}$ $10^{-6}$ $10^{-6}$ $10^{-6}${]} \\
%                   & $R_2$  &{[}800 800{]} \\
\hline
\end{tabular}
\label{table}
\end{table}

\subsection{Trajectory Tracking}\label{tt}

% -----------------------------------------------

\begin{figure*}[htbp]
\centering
\subfigure[]
{
    \begin{minipage}[b]{.3\linewidth}
        \centering
        \includegraphics[scale=0.29]{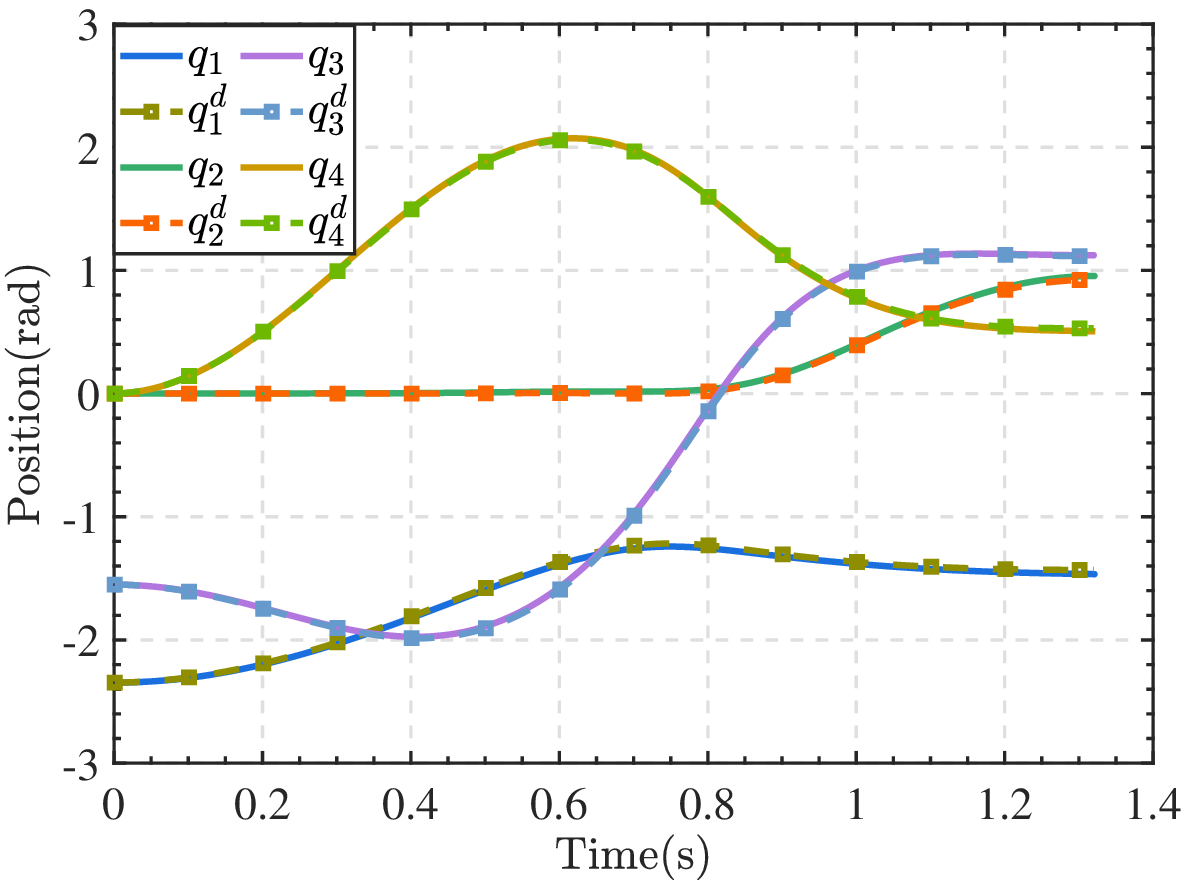}
    \end{minipage}
    \label{MPC_joint1}
}
\subfigure[]
{
 	\begin{minipage}[b]{.3\linewidth}
        \centering
        \includegraphics[scale=0.29]{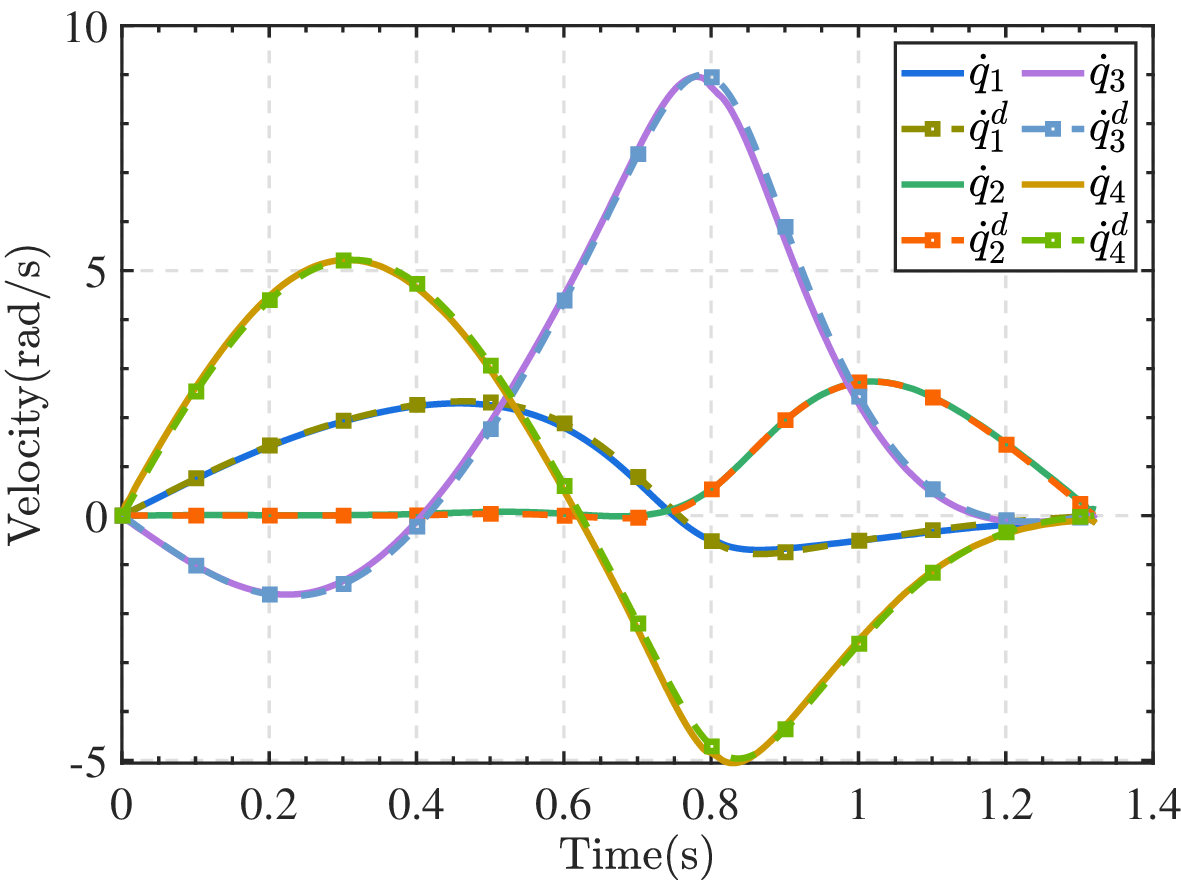}
    \end{minipage}
        \label{MPC_joint2}
}
\subfigure[]
{
 	\begin{minipage}[b]{.3\linewidth}
        \centering
        \includegraphics[scale=0.29]{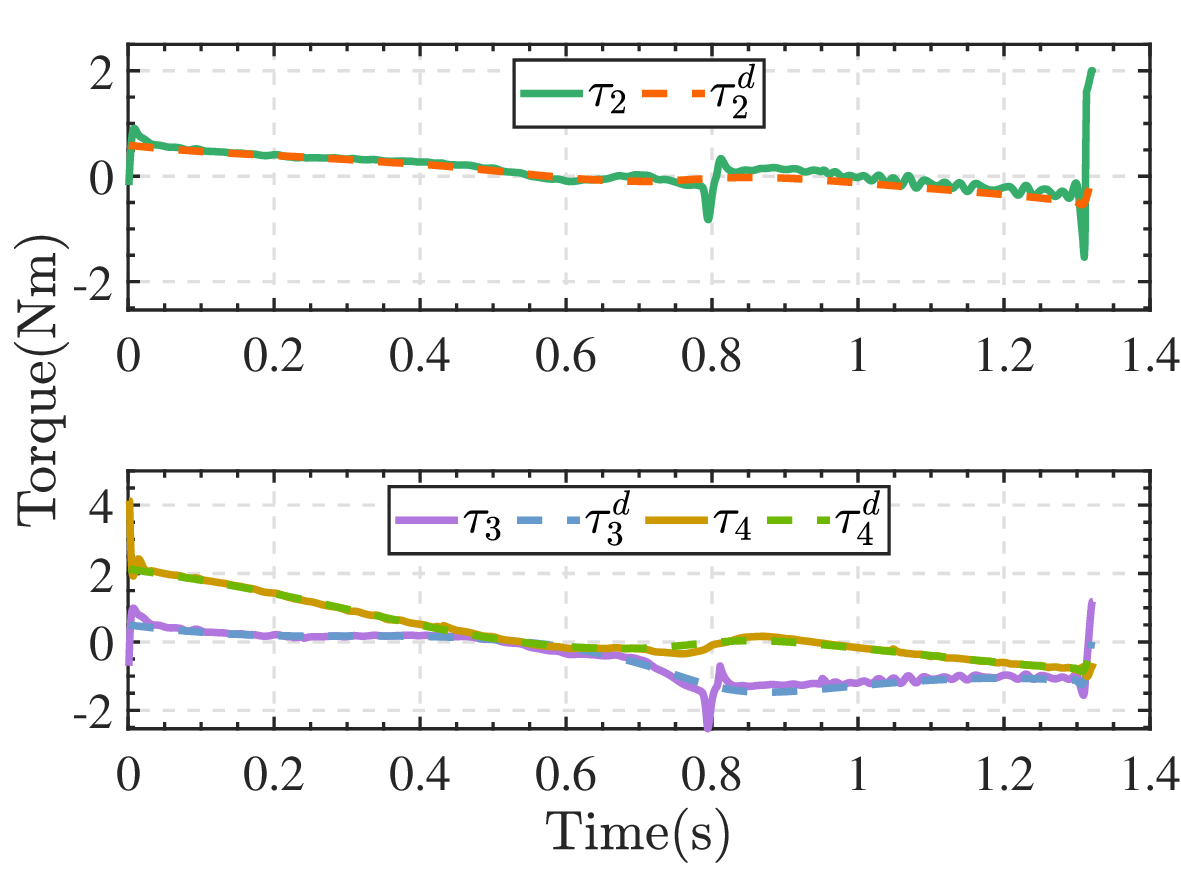}
    \end{minipage}
    \label{MPC_joint3}
}
\caption{\label{fig_MPC2}Trajectory tracking of MPC in joint space. (a), (b) and (c) show the tracking of joint position, joint velocity and joint torque, respectively.}
\label{MPC_joint}
\end{figure*}

\begin{figure}[htbp]
\subfigure[]
{
    \begin{minipage}[b]{.9 \linewidth}
        \centering
        \includegraphics[scale=0.29]{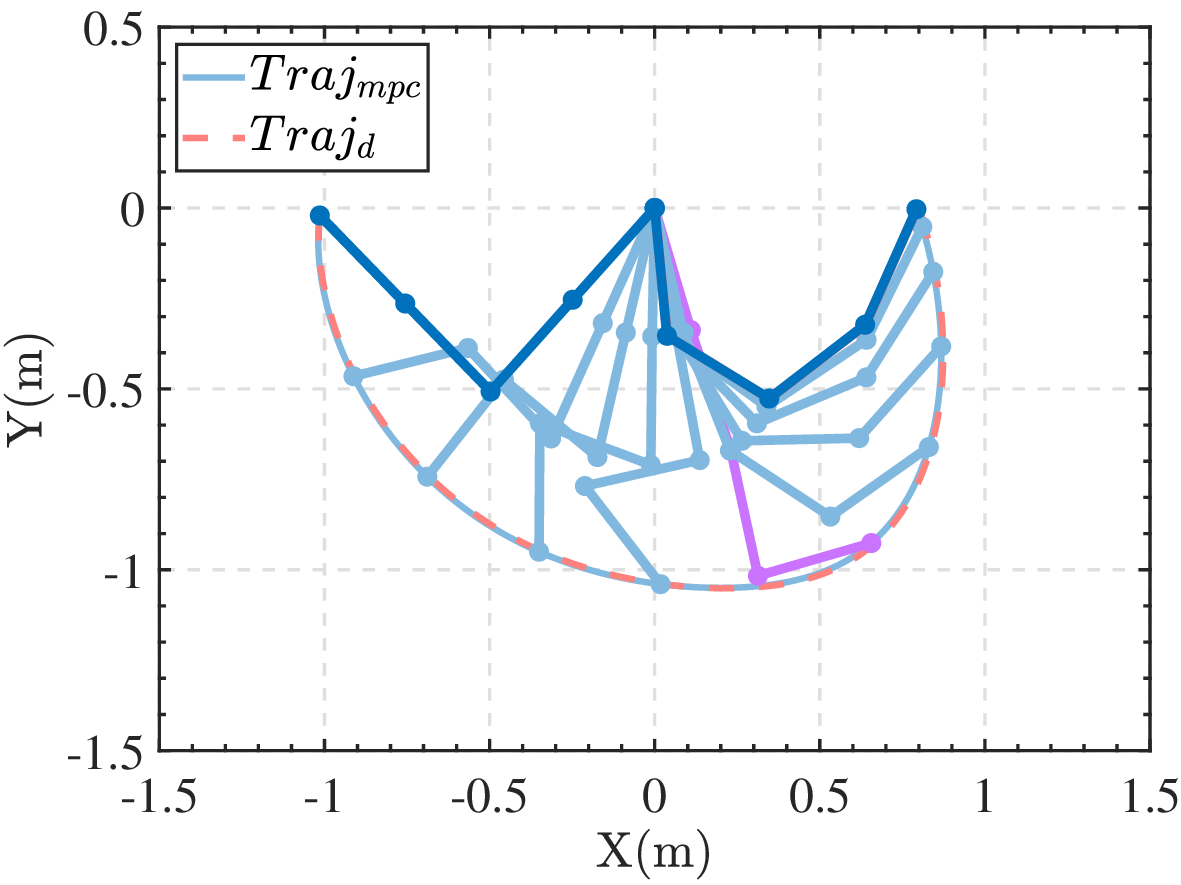}
    \end{minipage}
    \label{MPC_motion}
}
\subfigure[]
{
 	\begin{minipage}[b]{.9 \linewidth}
        \centering
        \includegraphics[scale=0.29]{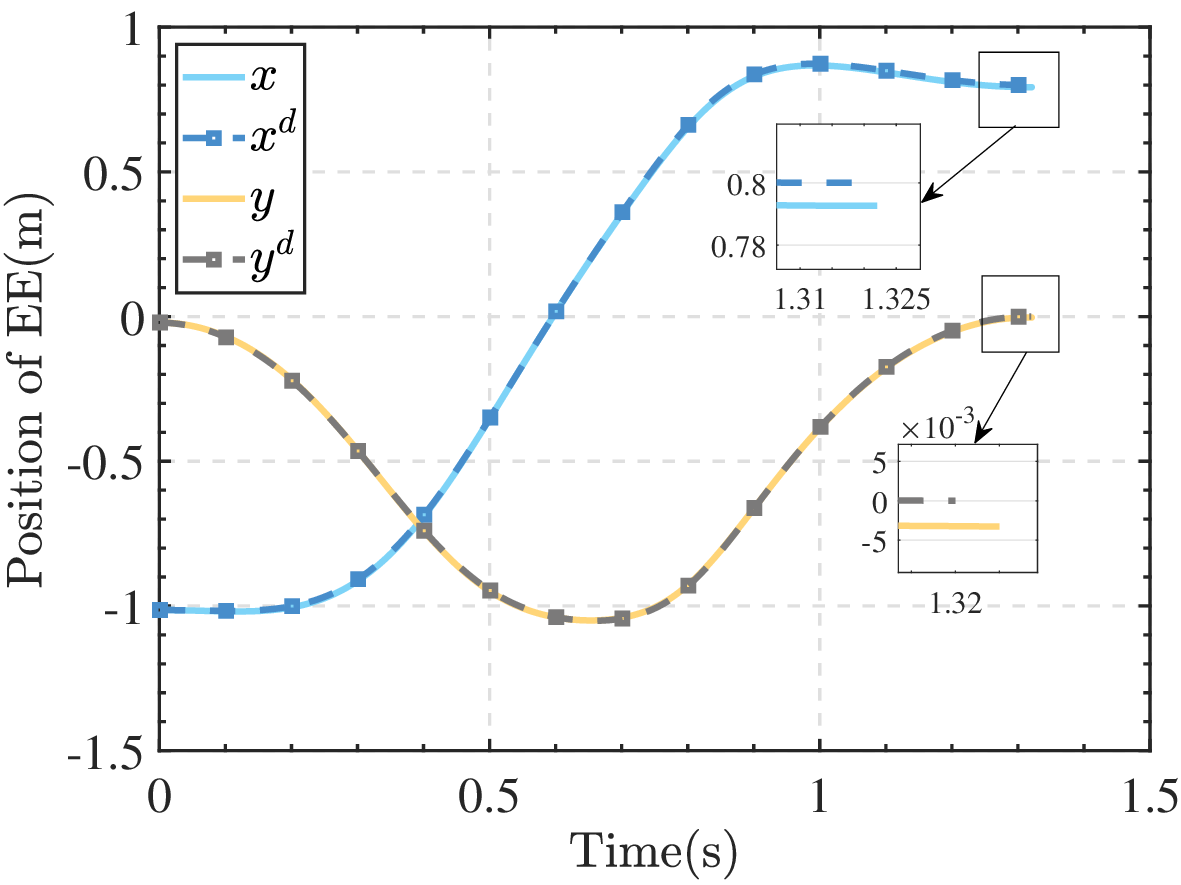}
    \end{minipage}
    \label{MPC_ee}
}
\caption{\label{fig_MPC1} The motion of the robot with MPC controller. (a) shows the stick diagram of the swing; (b) shows the EE tracking.}
\end{figure}

We initiate the robot's motion from the stationary position $\left[ q_1,q_2,q_3,q_4 \right] ^T=\left[ -2.35,0,-1.55,0 \right] ^T$, where the start point is (-1m,0m). The target point is (0.8m,0). The motion process is depicted in Fig.~\ref{fig_MPC1}, while the joint data is presented in Fig.~\ref{fig_MPC2}.

From Fig.~\ref{MPC_ee}, it can be inferred that the EE position exhibits an error of less than 0.02m in both the x and y directions at the final time. Considering the total length of the robot, this error can be considered acceptable, because an appropriately sized gripper can compensate for this deviation.
It can be clearly observed that the tracking performance of each joint is excellent from Fig.~\ref{fig_MPC2}.
Notably, as depicted in Fig.~\ref{MPC_joint1}, the underactuated joint 1 deviates slightly below the desired trajectory after 0.8s. Consequently, joint 2 surpasses slightly the desired trajectory after 0.8s to rectify the EE position error. In Fig.~\ref{MPC_joint3}, slight fluctuations are observed in the torque of joints 2 and 3 at 0.8s and 1.3s. The former is because at 0.8s, the robot is located in the purple position shown in Fig.~\ref{MPC_motion}, the first three links are straightened, resulting in singular position. The latter is due to the deceleration when approaching the target velocity. It is worth highlighting that the generated torques throughout the entire motion remain within the acceptable range of $\left[ -2.5N, +4N \right]$. The total energy consumption of the desired trajectory is 5.27J and that of the MPC trajectory is 5.31J, which is not much different. In summary, the presented figures demonstrate the effective tracking performance of the controller, thereby affirming the efficacy of our MPC controller.

% -----------------------------------------------
% \begin{figure}[H]
% \subfigure[]
% {
%     \begin{minipage}[b]{.459 \linewidth}
%         \raggedleft
%         \includegraphics[scale=0.217]{EXP2_fig1_a.eps}
%     \end{minipage}
%     \label{MPC_motion}
% }
% \subfigure[]
% {
%  	\begin{minipage}[b]{.459 \linewidth}
%         \raggedright
%         \includegraphics[scale=0.217]{EXP2_fig1_b.eps}
%     \end{minipage}
%     \label{MPC_ee}
% }
% \caption{\label{fig_MPC1} The motion of the robot in the MPC controller. (a) shows the schematic diagram of the swing, (b) shows the EE tracking.}
% \end{figure}
% -----------------------------------------------

\subsection{Obstacle Avoidance}\label{oae}
\begin{figure}[htbp]
\flushleft
\subfigure[]
{
    \begin{minipage}[b]{.9 \linewidth}
        \centering
        \includegraphics[scale=0.29]{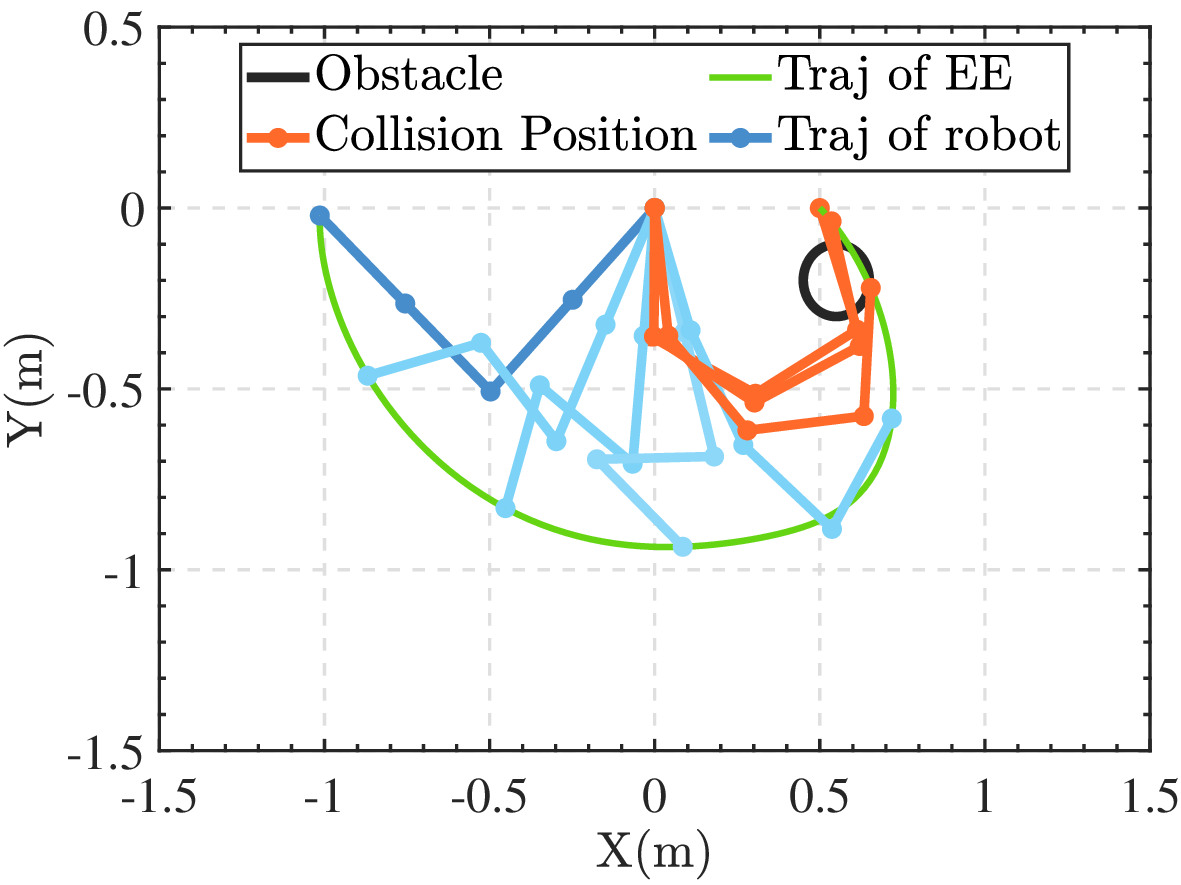}
    \end{minipage}
    \label{motion_no_oa}
}
\subfigure[]
{
 	\begin{minipage}[b]{.9 \linewidth}
        \centering
        \includegraphics[scale=0.29]{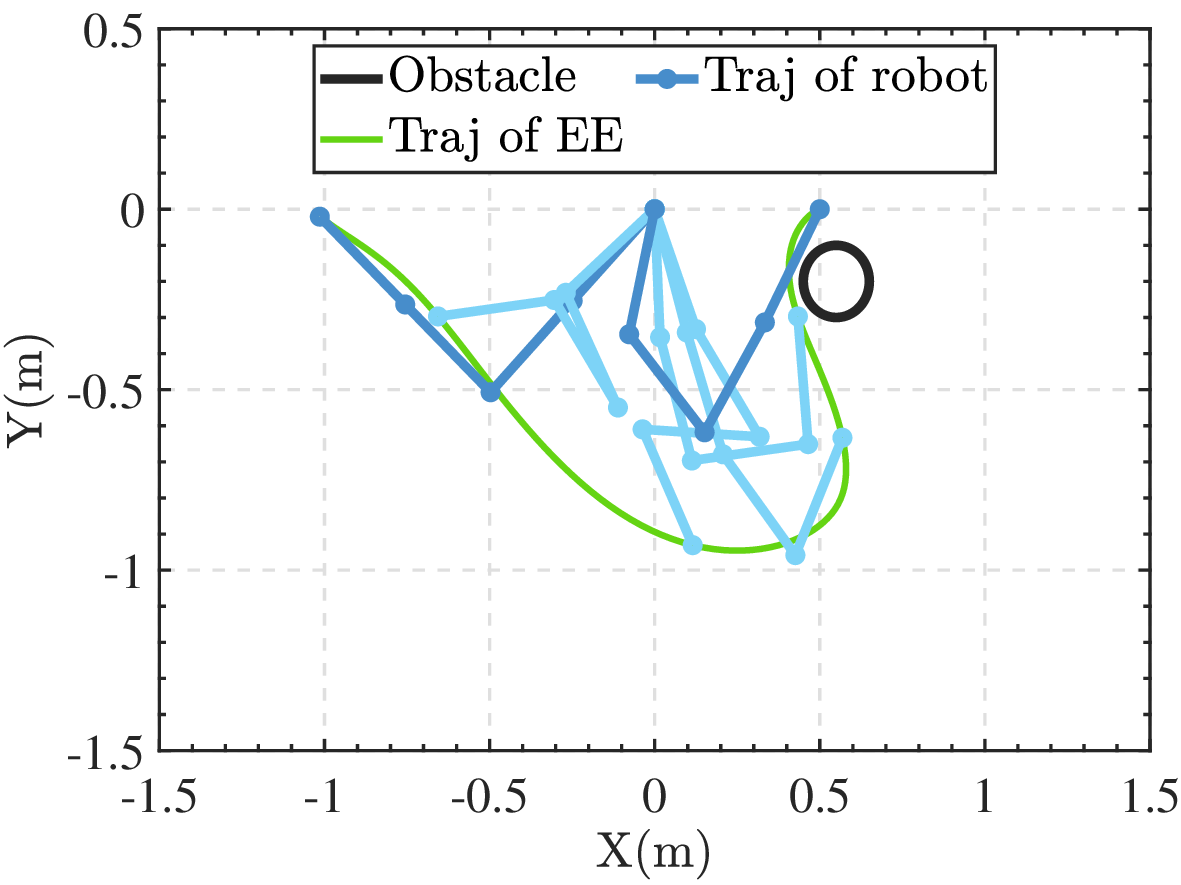}
    \end{minipage}
    \label{motion_oa}
}

\caption{\label{oa_test}The movement of the robot when there are obstacles. (a) shows the movement without the obstacle avoidance strategy; (b) shows the movement with the obstacle avoidance strategy.}
\end{figure}

The position of the obstacle can be set at any position on the path of the robot's swinging motion, here we select an example with an obstacle radius of 0.1m at coordinates (0.55m, -0.2m) for display. This obstacle is present throughout the movement. The robot's initial state is same as section \ref{tt}, and the target point is set at (0.5m, 0m). 
In Fig.~\ref{motion_no_oa}, it can be observed that the robot's swing end makes contact with the obstacle. However, when the obstacle avoidance strategy is employed, the robot's trajectory undergoes slight adjustments while still preserving the overall ``arcing" pattern. Specifically, the trajectory remains unchanged during the initial swing phase. The arms near the grip point and swing end contract and extend, respectively, during the later phase.
%------------------------------------------------
% \begin{figure}[htbp]
% \flushleft
% \subfigure[]
% {
%     \begin{minipage}[b]{.459 \linewidth}
%         \raggedleft
%         \includegraphics[scale=0.217]{EXP3_fig1_a.eps}
%     \end{minipage}
%     \label{motion_no_oa}
% }
% \subfigure[]
% {
%  	\begin{minipage}[b]{.459 \linewidth}
%         \raggedright
%         \includegraphics[scale=0.217]{EXP3_fig1_b.eps}
%     \end{minipage}
%     \label{motion_oa}
% }

% \caption{\label{oa_test}The movement of the robot when there are obstacles. (a) shows the movement without the obstacle avoidance strategy, (b) shows the movement with the obstacle avoidance strategy.}
% \end{figure}
%------------------------------------------------
%------------------------------------------------
\subsection{Robustness Verification}\label{rt}
\begin{figure}[htbp]
\flushleft
\subfigure[]
{
    \begin{minipage}[b]{.9 \linewidth}
        \centering
        \includegraphics[scale=0.29]{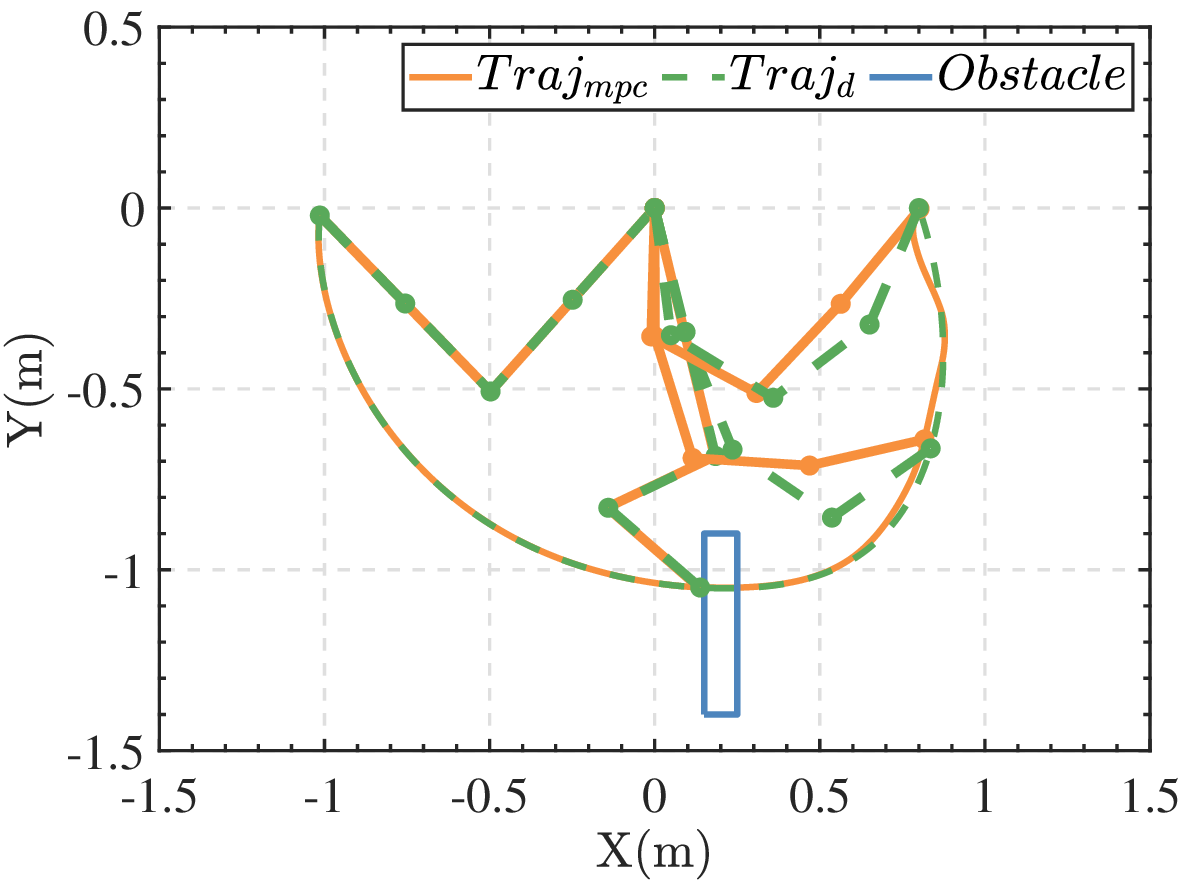}
    \end{minipage}
    \label{robusness_motion}
}
\subfigure[]
{
 	\begin{minipage}[b]{.9 \linewidth}
        \centering
        \includegraphics[scale=0.29]{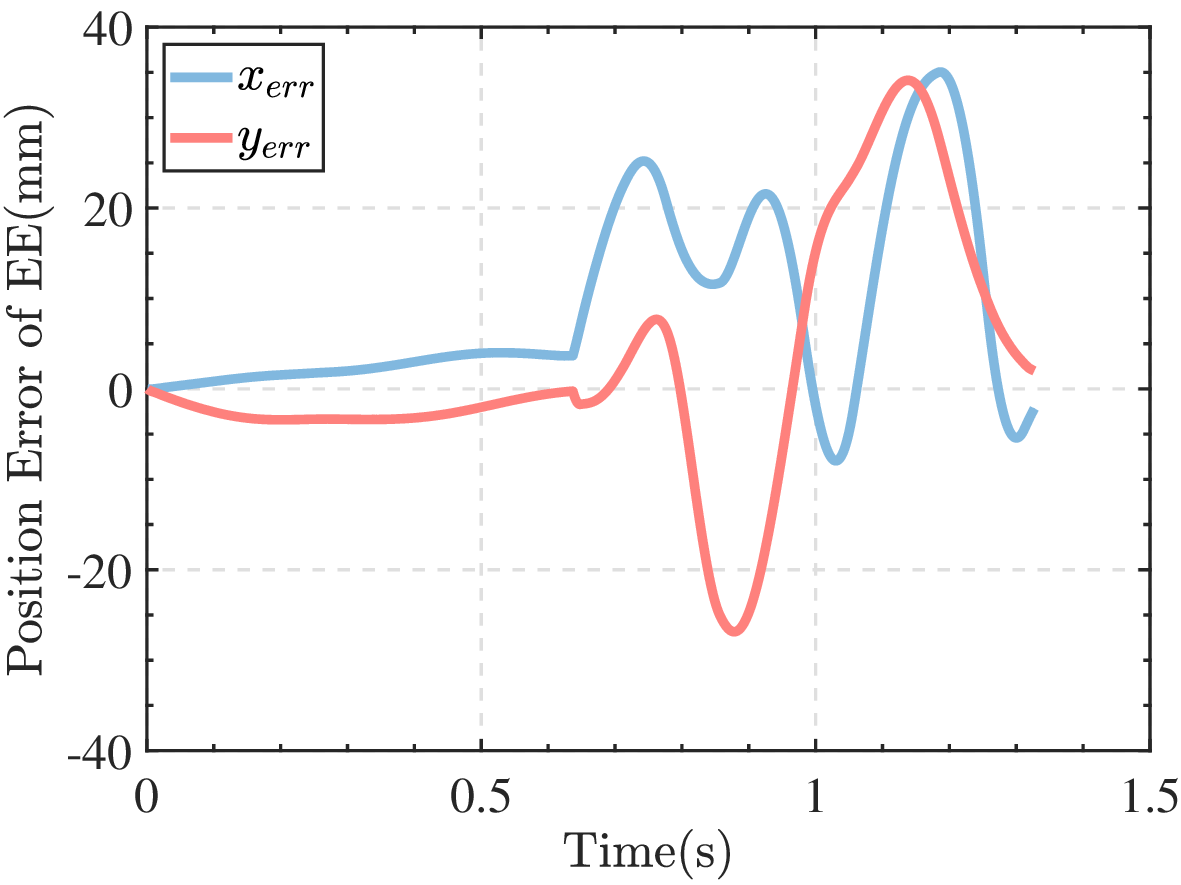}
    \end{minipage}
    \label{robusness_err}
}

\caption{\label{robusness}The movement when the robot collides with an obstacle. (a) shows the stick diagram of motion; (b) shows the EE errors.}
\end{figure}

To assess the controller's robustness during the swing process, we introduced an obstacle along the trajectory. When the robot hits the obstacle without avoidance, it knocks the obstacle away. The robot's initial state and target point are same as section \ref{tt}. The obstacle is positioned at (-1.2m, 0.2m) and had a mass of 0.4kg. 
% Fig.~\ref{robusness} presents the motion trajectory of the robot during the collision process, along with the EE error.
As depicted in Fig.~\ref{robusness}, the collision occurred at 0.62s, resulting in an increase in the EE error. The trajectory tracking in joint space exhibited significant deviations, particularly for the underactuated joints, shwon in Fig.\ref{robusness_motion}. However, thanks to the incorporation of EE error weighting in the MPC framework, the robot's EE trajectory remained relatively close to the desired trajectory, and the error diminished to nearly zero finally, shown in Fig.\ref{robusness_err}. 

%------------------------------------------------
% \begin{figure}[htbp]
% \flushleft
% \subfigure[]
% {
%     \begin{minipage}[b]{.459 \linewidth}
%         \raggedleft
%         \includegraphics[scale=0.217]{EXP4_fig1_a.eps}
%     \end{minipage}
%     \label{robusness_motion}
% }
% \subfigure[]
% {
%  	\begin{minipage}[b]{.459 \linewidth}
%         \raggedright
%         \includegraphics[scale=0.217]{EXP4_fig1_b.eps}
%     \end{minipage}
%     \label{robusness_err}
% }

% \caption{\label{robusness}Motion tracking when the robot collides with an obstacle. (a) shows the schematic diagram of the motion, and (b) shows the EE errors.}
% \end{figure}
%------------------------------------------------
%------------------------------------------------
\section{Conclusions And Future Work}
\vspace{-0.75mm}
This paper presents dynamics modeling and trajectory generation and tracking methods for a four-link underactuated brachiation robot. An offline trajectory generation method based on energy optimization is developed by the direct collocation to generate joint-space trajectories. Subsequently, a linear Model Predictive Control controller is employed for online trajectory tracking in both joint space and task space. The simulation results prove that the trajectory generation and tracking methods are effective. We also compare the total energy consumption under different lower-to-upper arm length ratios and swing times, leading to the selection of the most suitable arm length ratio and validating the reasonableness of the swing time setting. The simulation results also demonstrate the robot has satisfied obstacle avoidance capability and robustness.

In future, we plan to focus on the hardware implementation of the robot. In terms of control strategy, we aim to explore the state-of-the-art reinforcement learning methods and compare them with the approach presented in this paper.
% \section*{Acknowledgment}

% The preferred spelling of the word ``acknowledgment'' in America is without 
% an ``e'' after the ``g''. Avoid the stilted expression ``one of us (R. B. 
% G.) thanks $\ldots$''. Instead, try ``R. B. G. thanks$\ldots$''. Put sponsor 
% acknowledgments in the unnumbered footnote on the first page.

% \section*{References}
% Please number citations consecutively within brackets \cite{b1}. The 
% sentence punctuation follows the bracket \cite{b2}. Refer simply to the reference 
% number, as in \cite{b3}---do not use ``Ref. \cite{b3}'' or ``reference \cite{b3}'' except at 
% the beginning of a sentence: ``Reference \cite{b3} was the first $\ldots$''

% Number footnotes separately in superscripts. Place the actual footnote at 
% the bottom of the column in which it was cited. Do not put footnotes in the 
% abstract or reference list. Use letters for table footnotes.

% Unless there are six authors or more give all authors' names; do not use 
% ``et al.''. Papers that have not been published, even if they have been 
% submitted for publication, should be cited as ``unpublished'' \cite{b4}. Papers 
% that have been accepted for publication should be cited as ``in press'' \cite{b5}. 
% Capitalize only the first word in a paper title, except for proper nouns and 
% element symbols.

% For papers published in translation journals, please give the English 
% citation first, followed by the original foreign-language citation \cite{b6}.

\vspace{12pt}
% \color{red}
% IEEE conference templates contain guidance text for composing and formatting conference papers. Please ensure that all template text is removed from your conference paper prior to submission to the conference. Failure to remove the template text from your paper may result in your paper not being published.

\end{document}